\def\hbar{{\mathchar'26\mkern-9mu h}}
\documentstyle[12pt]{article}
\advance\textwidth 1in
\advance\oddsidemargin -0.5in
\advance\textheight 1in
\advance\topmargin -0.5in
\title{Eigenvalue Integro-Differential Equations for Orthogonal Polynomials
on the Real Line}

\author{\\ \\Carl M. Bender\\
        Physics Department\\
        Washington University\\
        St. Louis, MO \ \ 63130 USA \\
		e-mail: cmb@howdy.wustl.edu\\
        \\
     Joshua Feinberg\\
	 Theory Group\\
     Department of Physics\\
     University of Texas at Austin\\
     Austin, TX \ \ 78712 USA \\
	 e-mail: joshua@utaphy.ph.utexas.edu\\}
\date{}
\begin{document}
\maketitle
\def\thepage{UTTG-15-94, WU-HEP-1994-17, hep-th/9411040}
\thispagestyle{myheadings}
\bigskip
\bigskip
\baselineskip=24pt
\def\cp{{\cal P}}
\def\cq{{\cal Q}}
\begin{abstract}
The one-dimensional harmonic oscillator wave functions are solutions to a
Sturm-Liouville problem posed on the whole real line. This problem generates
the
Hermite polynomials. However, no other set of orthogonal polynomials can be
obtained from a Sturm-Liouville problem on the whole real line. In this paper
we
show how to characterize an arbitrary set of polynomials orthogonal on
$(-\infty,\infty)$ in terms of a system of integro-differential equations of
Hartree-Fock type. This system replaces and generalizes the linear differential
equation associated with a Sturm-Liouville problem. We demonstrate our results
for the special case of Hahn-Meixner polynomials.
\end{abstract}
\newpage
\baselineskip=24pt
\pagenumbering{arabic}
\addtocounter{page}{1}

\section{Introduction}

In this paper we will examine the system of $N$ coupled nonlinear
integro-differential equations defined on the whole real line
\begin{eqnarray}
&&\left[-{1\over 2}{d^2\over dx^2}-{1\over 4}U^{\prime\prime}(x)+{1\over 8}
U^{\prime^2}(x)+\mu_i^{(N)}\right]u_i(x)-{1\over 2}\sum_{j=0}^{N-1}
\int_{-\infty}^{\infty} dy\, u^*_j(y)
\nonumber \\
&&\quad\times\frac{U^\prime(x)
-U^\prime(y)}{x-y}\left[u_i(x)u_j(y)-u_j(x)u_i(y)
\right] \ = 0\quad (i=0,\,1,\,2,\,\ldots,\,N-1).
\label{1}
\end{eqnarray}
This system is an eigenvalue problem in which there are $N$ eigenfunctions
$u_i(x)$ and $N$ corresponding eigenvalues $\mu_i^{(N)}$. The eigenfunctions
are
constrained by the orthonormality condition
\begin{equation}
\int_{-\infty}^{\infty} dx\,u_i (x) u_j(x) = \delta_{ij}.
\label{2}
\end{equation}
The single function $U(x)$ together with the integer $N$ completely
characterize
the eigenvalue problem under consideration. The function $U(x)$ is restricted
by
the requirements that it be twice continuously differentiable, that it be
bounded from below, and that it satisfy the inequality constraint
\begin{equation}
\int_{-\infty}^{\infty}dx\,e^{-U(x)}<\infty .
\label{3}
\end{equation}

This eigenvalue problem constitutes a system of Hartree-Fock equations that
describe the ground state of a one-dimensional Fermi gas having $N$ particles.
The functions $u_i(x)$ are the single-particle states of the Fermi gas and the
numbers $\mu_i^{(N)}$ are the corresponding chemical potentials associated with
(\ref{2}).

As will be shown later, (\ref{1}-\ref{3}) arise in the study of random
Hermitian
$N$-dimensional matrices $\Phi$ whose statistical weight is given by
\cite{Mehta}
\begin{equation}
d\mu(\Phi)={1\over Z}e^{-{\rm Tr}\, U(\Phi)} d^{N^2}\Phi,
\label{4}
\end{equation}
where $Z$ is a normalization factor \cite{Josh}. Associated with such an
ensemble of random matrices is a set of polynomials ${\cal P}_i(x)$ that are
orthonormal on the whole real line with respect to the weight in (\ref{4})
\cite{BIZ}:
\begin{equation}
\int_{-\infty}^{\infty}dx\, {\cal P}_i(x){\cal P}_j(x)e^{-U(x)}=\delta_{ij}.
\label{5}
\end{equation}
The relation between the polynomials $\{{\cal P}_i(x)\}$ and the eigenfunctions
$\{u_i(x)\}$ of (\ref{1}) is simply
\begin{equation}
u_i(x)=e^{-{1\over 2}U(x)}{\cal P}_i(x).
\label{6}
\end{equation}
If we substitute (\ref{6}) into (\ref{1}) and use (\ref{2}) we can calculate
the eigenvalues $\{\mu_i^{(N)}\}$. The connection between the eigenvalue
problem
(\ref{1}) and the theory of random matrices and orthogonal polynomials will be
discussed in more detail in Sec.~III. However, the principal consequence of
this
connection is that (\ref{6}) is the {\sl unique\/} solution
$[u_0(x),\,u_1(x),\,
u_2(x),\,\ldots, u_{N-1}(x)]$ to (\ref{1}) and (\ref{2}) for a given function
$U(x)$. Physically, this uniqueness is a consequence of the Fermi gas being in
its ground state. To the best of our knowledge, the eigenvalue problem in
(\ref{1}) is new in the general theory of orthogonal polynomials.

In the special case $U(x)={1\over 2}x^2$, (\ref{1}) simplifies to a set of $N$
decoupled linear differential equations:
\begin{equation}
\left( -{d^2\over dx^2}+{1\over 4}x^2+2\mu_i^{(N)}-N+{1\over 2}\right)u_i(x)=0.
\label{7}
\end{equation}
These are the Schr{\"o}dinger equations for the lowest $N$ eigenstates of the
one-dimensional harmonic oscillator from which we immediately infer that
$\mu_i^{(N)}={1\over 2}(N-1-i)$. In the case of the harmonic oscillator,
$\mu_i^{(N)}$ is the depth of the $i$th state below the Fermi level.

We will now show that the Hermite polynomials are the {\sl only\/} set of
orthogonal polynomials that are determined by a Sturm-Liouville problem {\sl on
the whole real line}. Let $\{{\cal P}_i(x)\}$ be a set of orthogonal
polynomials
satisfying (\ref{5}). Assume further that the functions $\{u_i(x)\}$ as defined
in (\ref{6}) are the eigenfunctions of the general Sturm-Liouville problem
\cite{Ince}
\begin{equation}
-{d\over dx}\left[ p(x){d\over dx}u_i(x)\right]+[q(x)-\mu_i W(x)]u_i(x)=0\quad
(-\infty<x<\infty),
\label{8}
\end{equation}
where $p(x)>0$ and $q(x)$ are continuous functions and $\{\mu_i\}$ are the
associated eigenvalues. Since the domain of (\ref{8}) is infinite, the only
boundary condition imposed on the eigenfunctions $\{ u_i(x)\}$ is the
normalization condition (\ref{2}). We emphasize that $p(x)$ is strictly
positive
for all $x$ on the whole line.\footnote{In our case the domain of definition of
the Sturm-Liouville problem is the whole real line, $-\infty<x<\infty$. In
cases
where the domain of definition is a compact subset of the real line, $p(x)$ may
vanish at its boundary points, which then become singular points of the
Sturm-Liouville problem. In the case considered here the function $p(x)$ may
vanish only at $x=\pm\infty$.} Moreover, since the Sturm-Liouville problem in
(\ref{8}) is one dimensional, the orthonormality conditions (\ref{5}) imply
that
the spectrum $\{\mu_i\}$ is discrete, nondegenerate, and bounded from below.
 Following (\ref{2}) we note further that the weight in (\ref{8}) is fixed by
$W(x)=1$. Substituting the functions $\{u_i(x)\}$ from (\ref{6}) into the
eigenvalue problem (\ref{8}), we obtain
\begin{equation}
p(x)\cp_i^{\prime\prime}(x)+\left[ 2{w^{\prime}(x)\over w(x)}p(x)+ p^{\prime}
(x)\right]\cp_i^{\prime}(x)+\left[ p(x) {w^{\prime\prime}(x) \over w(x)}+
p^{\prime}(x){w^{\prime}(x) \over w(x)}+\mu_i-q(x)\right]\cp_i(x)=0,
\label{9}
\end{equation}
where $w(x)=e^{-{1\over 2} U(x)}$. Equation (\ref{9}) holds for all real $x$
and may be used to express $p(x),~q(x),~w(x)$, and the spectrum $\{\mu_i\}_{i
\geq 3}$ in terms of the first three polynomials and their corresponding
eigenvalues as we now show.

Let the first three orthonormal polynomials in (\ref{6}) be
\begin{equation}
\cp_0(x)=1,\quad\quad\cp_1(x)=ax+b\quad (a\neq 0),\quad\quad
\cp_2(x)=cx^2+dx+e\quad (c\neq 0),
\label{10}
\end{equation}
where we have assumed the normalization
\begin{equation}
\int_{-\infty}^{\infty}~dx~w^2(x)~=~1.
\label{11}
\end{equation}
Substituting $\cp_0~{\rm and}~\cp_1$ from (\ref{10}) into (\ref{9}) we obtain
\begin{eqnarray}
&&q(x)=p(x){w^{\prime\prime}(x)\over w(x)}+p^{\prime}(x){w^{\prime}(x)\over
w(x)}+\mu_0,\nonumber \\
&&\left[ 2{w^{\prime}(x)\over w(x)}p(x)+ p^{\prime}(x)\right]+\left(\mu_1-\mu_0
\right)\left(x+{b \over a}\right)=0.
\label{12}
\end{eqnarray}
These results allow us to rewrite (\ref{9}) as
\begin{equation}
p(x)~\cp_i^{\prime\prime}(x)-\left(\mu_1-\mu_0\right)\left(x+{b \over a}\right)
\cp_i^{\prime}(x)+\left(\mu_i-\mu_0\right)\cp_i(x)=0.
\label{13}
\end{equation}
Substituting $\cp_2 (x)$ from (\ref{10}) into (\ref{13}) we find that
\begin{eqnarray}
p(x)&=&\left(\mu_1-\mu_0-{\mu_2-\mu_0 \over
2}\right)x^2+\left[\left(\mu_1-\mu_0
\right){b \over a}-\left(\mu_2-\mu_1\right){d \over 2c}\right]x\nonumber\\
&&\quad\quad +{1\over 2c}\left[\left(\mu_1-\mu_0\right){bd\over a}-
\left(\mu_2-\mu_0\right)e\right].
\label{14}
\end{eqnarray}

The function $p(x)$ is strictly positive for any real value of $x$. Thus,
either
$p(x)$ is a quadratic polynomial and the requirement of positivity implies the
two inequalities
\begin{eqnarray}
\mu_2+\mu_0-2\mu_1 &<& 0,\nonumber\\
\nonumber\\
\left(\mu_1-\mu_0\right)^2\left({b\over a}-{d\over 2c}\right)^2+\left(\mu_2-
\mu_0\right)\left(\mu_2+\mu_0-2\mu_1\right)
\left[\left({d \over 2c}\right)^2-{ e \over c}\right] &<& 0,
\label{15}
\end{eqnarray}
or else $p(x)$ is a positive constant and we have
\begin{eqnarray}
\mu_1-\mu_0-{\mu_2-\mu_0 \over 2}&=&0,\nonumber\\
\left(\mu_1-\mu_0\right){b\over a}-\left(\mu_2-\mu_1\right){d\over 2c}&=&0,
\nonumber\\
{1\over 2c}\left[\left(\mu_1-\mu_0\right){bd \over a}-\left(\mu_2-\mu_0
\right)e\right] &>& 0.
\label{16}
\end{eqnarray}

Equation (\ref{14}) is only one of many possible expressions for $p(x)$.
Indeed,
one may alternatively solve (\ref{13}) for $p(x)$ in terms of any one of the
polynomials $\{\cp_i\}_{i\geq 3}$ as
\begin{equation}
p(x)={\left(x+{b\over a}\right)\left(\mu_1-\mu_0\right)\cp_i^{\prime}(x)-
\left(\mu_i-\mu_0\right)\cp_i(x)\over \cp_i^{\prime\prime}(x)},
\label{17}
\end{equation}
which owing to (\ref{14}), must be either a quadratic polynomial in $x$ or a
positive constant for all $i$. This is precisely the condition that determines
the whole spectrum in terms of $\mu_0,~\mu_1,~{\rm and}~\mu_2~ $. To see this,
we take the limit of (\ref{17}) as $|x|\rightarrow\infty$ and using (\ref{14})
we find that
\begin{equation}
\mu_i=\left[{\mu_2-\mu_0 \over 2}-\left(\mu_1-\mu_0\right)\right] i^2+
\left[2\left(\mu_1-\mu_0\right)-{\mu_2-\mu_0 \over 2}\right] i +\mu_0.
\label{18}
\end{equation}

The condition that the spectrum in (\ref{18}) increases monotonically implies
the inequality
\begin{equation}
\mu_2+\mu_0-2\mu_1\geq 0.
\label{19}
\end{equation}
If $p(x)$ is a quadratic polynomial, then the inequality $\mu_2+\mu_0-2\mu_1>0$
must hold, but this contradicts the first equation in (\ref{15}). It also
contradicts the second inequality in (\ref{15}) because if both these relations
are valid, they imply the inequality
\begin{equation}
\left({d \over 2c}\right)^2-{e\over c}<0.
\label{20}
\end{equation}
However, from (\ref{10}) this is precisely the condition that $\cp_2(x)$ be
nonvanishing on the real line, which implies that its sign never alternates
along the real line. This conclusion contradicts the orthogonality of
$\cp_2~{\rm and}~\cp_0$.\footnote{There is another way to see that (\ref{20})
is
false. We note that the coefficient $e$ in $\cp_2(x)$ in (\ref{10}) is actually
not arbitrary; it may be determined in terms of the coefficients $a$, $b$, $c$,
and $d$ as follows. We consider the following three orthonormality relations:
(1) $\int dx\,w^2(x)\cp_0(x)\cp_1(x)=0$; (2) $\int dx\,w^2(x)\cp_0(x)\cp_2(x)
=0$; (3) $\int dx\,w^2(x)\cp_1^2(x)=1$. These equations may be regarded as
three
linear simultaneous equations for three unknowns, the first and second moments
of $w^2(x)$ [the zeroth moment is given in (\ref{11})] and the coefficient $e$.
The solution for $e$ is $e=(abd-c-b^2c)/a^2$. When this value is substituted
into the inequality in (\ref{20}), the inequality takes the form $[(ad-2bc)^2
+4c^2]/(4a^2c^2)<0$, which is manifestly impossible.} Therefore, (\ref{15}) can
never hold, and (\ref{16}), which implies that $p(x)$ is a positive constant
$p$, is the only possibility. Finally, using (\ref{11}-\ref{12}) we find that
\begin{equation}
w^2(x)=e^{-U(x)}=\left({2\pi p\over \mu_1-\mu_0}\right)^{1/2}
e^{-{\mu_1-\mu_0\over 2p}\left(x+{b\over a}\right)^2}
\label{21}
\end{equation}
and
\begin{equation}
q(x)={\left(\mu_1-\mu_0\right)^2\over 4p}\left(x+{b\over a}\right)^2
+{3\over 2}\mu_0 - {1\over 2}\mu_1,
\label{22}
\end{equation}
for which (\ref{8}) becomes the Schr{\"o}dinger equation for an harmonic
oscillator whose equilibrium point is at $x_0=-b/a$, and the polynomials
$\{\cp_i(x)\}$ are the corresponding (shifted) Hermite polynomials and the
eigenvalues are $\mu_i=(\mu_1-\mu_0)i+\mu_0$. This concludes the demonstration
that the Hermite polynomials are the only set of orthogonal polynomials that
are
determined by a Sturm-Liouville problem on the whole real line.

This paper is organized as follows. In Sec.~II we present a proof of the
integro-differential eigenvalue equation (\ref{1}) based on a variational
calculation. In Sec.~III we give a general formula for the eigenvalues
associated with an arbitrary system of polynomials having an even weight
function. This formula for the eigenvalues is useful because it gives the
spectrum for a one-dimensional system of interacting fermions. Finally, in
Sec.~IV we illustrate equations (\ref{1}) and (\ref{2}) for the special case of
Hahn-Meixner polynomials. We calculate the first few eigenvalues and find some
of their general features.
\pagebreak

\section{Orthonormal Polynomials on the Real Line}

In this section we use a variational principle to prove that polynomials
orthonormal relative to an arbitrary weight $w^2(x)=e^{-U(x)}$ on the whole
real
line must obey the system of Hartree-Fock equations in (\ref{1}) \cite{Josh}.
We show that as a consequence the functions in (\ref{6}) are the {\sl unique\/}
solution of (\ref{1}). [Recall from Sec.~I that the system of equations in
(\ref{1}) reduces to the trivial set of uncoupled Schr\"{o}dinger equations
(\ref{7}) for the eigenstates of an harmonic oscillator (i.e., the case of
Hermite polynomials) when the weight is Gaussian $w^2(x)=e^{-x^2/2}$.]

Consider a quantum mechanical system whose degrees of freedom are elements of
an
$N\times N$ Hermitian matrix $\Phi$. The Hamiltonian for this system is the
positive semi-definite operator given by
\begin{equation}
{\cal H}=\frac{1}{2}{\rm Tr}\left[\left( -\frac{\partial}{\partial\Phi}+
\frac{1}{2}U^{\prime}(\Phi)\right)\left(\frac{\partial}{\partial\Phi}+
\frac{1}{2} U^{\prime} (\Phi)\right)\right].
\label{3.1}
\end{equation}
We will require that $U(\Phi)$ be a matrix potential function satisfying the
restriction in (\ref{3}). This clearly implies that the potential in
(\ref{3.1})
is bounded from below and grows to plus infinity as the matrix eigenvalues
become infinite. Hence, the Schr\"{o}dinger operator ${\cal H}$ in (\ref{3.1})
has a well-defined spectrum. This Hamiltonian is symmetric under the adjoint
$U(N)$ transformation $\Phi\,\to\,\Omega\Phi\Omega^\dagger$. Therefore,
${\cal H}$ possesses a unique ($U(N)$ singlet) normalizable ground-state vector
$\Psi_0(\Phi)$. This ground state has energy zero and is given by\footnote{It
is
easy to verify that this state has zero energy by substituting (\ref{3.2}) into
(\ref{3.1}). Thus, since ${\cal H}$ is non-negative, $\Psi_0$ in (\ref{3.2}) is
the ground state because the latter is unique.}
\begin{eqnarray}
\Psi_0(\Phi)&=&{1\over\sqrt{Z}}\exp[-{1\over 2}{\rm Tr}U(\Phi)] \nonumber\\
Z&=&\int d^{N^2}\Phi\exp [-{\rm Tr}\;U(\Phi)].
\label{3.2}
\end{eqnarray}

As is well known, the Laplacian over Hermitian matrices acquires the form
\cite{BIZ,Onofri,BPIZ}
\begin{equation}
-{\rm Tr}\frac{\partial^2}{\partial\Phi^2}=-\frac{1}{\Delta(x)}\sum^N_{i=1}
\frac{\partial^2}{\partial x^2_i}\Delta (x)+[{\rm U}(N)~{\rm angular~momentum~
terms}],
\label{3.3}
\end{equation}
where $x_i$ are the matrix eigenvalues and $\Delta(x_i)$ is the Vandermonde
determinant.

 For a generic potential $U(\Phi)$, the Hamiltonian in (\ref{3.1}) contains
long-range two-body interaction terms between the Fermions\footnote{Note that
if $U$ is a polynomial of degree less than or equal to three, there are no
two-body interactions in (\ref{3.4}).}
\begin{equation}
{\cal H}_{\rm int}=-\frac{1}{4}{\rm Tr}\left[\left(\frac{\partial}{\partial
\Phi}\right) U^\prime(\Phi)\right]=-\frac{1}{4}\sum_{i,j}
\frac{U^\prime(x_i)-U^\prime(x_j)}{x_i-x_j}.
\label{3.4}
\end{equation}
Thus, a one-dimensional collection of eigenvalues may be considered as an
interacting Fermi gas.

The eigenstates of ${\cal H}$ have definite $U(N)$ quantum numbers and thus
fall
into definite $U(N)$ representations. Because of this symmetry it is useful to
introduce matrix polar coordinates $\Phi=\Omega X\Omega^\dagger$, where $\Omega
\in U(N)$ and $X$ is a real diagonal matrix whose entries, $x_1,\,x_2\,\ldots,
x_N$, are the eigenvalues of $\Phi$. The transformation from Cartesian matrix
coordinates $\Phi_{ij}$ to polar matrix coordinates $\Omega_{ij},\,{\rm and}\,
X_{ij}=x_i\delta_{ij}$ is associated with a Jacobian, which is proportional to
$\Delta^2(x_i)$. Under this transformation ${\cal H}$ and $\Psi_0$ transform as
\begin{eqnarray}
{\cal H}&\to&\Delta {\cal H}\Delta^{-1},\nonumber\\
\Psi_0 &\to&\Delta\Psi_0.
\label{3.5}
\end{eqnarray}
This implies that the transformed self-adjoint Hamiltonian is symmetric and
that
the transformed ground state is totally anti-symmetric under the interchange of
any two of the eigenvalues $x_i$. These symmetry properties remain valid for
any
$U(N)$ singlet eigenstate of ${\cal H}$. Therefore, the singlet sector of
(\ref{3.1}) is equivalent to a one-dimensional Fermi gas with a fixed number
$N$
of particles \cite{BPIZ}, where the eigenvalues $x_i$ are considered as the
coordinates of the $N$ Fermi particles.

Let us consider the class of singlet states of the form
\begin{equation}
\Psi_V(\Phi)=\exp [-{1\over 2}{\rm Tr}V(\Phi)],
\label{3.6}
\end{equation}
where $\Psi_V$ is assumed to be normalizable. In matrix polar coordinates this
becomes
\begin{equation}
\Psi_V(x_i)={\cal N}{\rm det}_{i,j}\,\left (x_j^{i-1}\right ){\rm exp}\left [ -
{1\over 2}\sum_{k=1}^N V(x_k)\right ]\quad (1\leq i,j\leq N),
\label{3.7}
\end{equation}
where we have used the identity $\Delta(x_j)={\rm det}_{i,j}\, (x_j^{i-1} )$
and
${\cal N}$ is a normalization constant.

It now becomes clear how this discussion of a Fermi gas relates to the theory
of
polynomials \cite{BIZ}. First, note that
\begin{equation}
{\rm det}_{i,j}\,\left (x_j^{i-1}\right )=
{\rm det}_{i,j}\, \left [ P_{i-1}(x_j) \right ]\quad (1\leq i,j\leq N)
\label{3.8}
\end{equation}
for any set of {\sl monic\/} polynomials $P_i(x)=x^i+({\rm
lower\,powers\,of}~x)
$. Thus, the wave function (\ref{3.7}) is a Slater determinant \cite{BJ}:
\begin{equation}
\Psi_V(x_i)={1\over\sqrt{N!}}{\rm det}_{i,j}\,\left [ v_{i-1}(x_j)\right ],
\label{3.9}
\end{equation}
where $v_i(x)=\nu_i P_i (x)e^{-V(x)/2}$ are interpreted as normalized single
particle states in the N fermion wave function (\ref{3.9}), which will be used
as a trial wave function in a variational (Hartree-Fock) calculation of the
ground state of the Fermi gas and ${\cal N}$ is a normalization constant.

Therefore, these functions must satisfy the orthonormality condition\cite{BJ}
\begin{equation}
\int_{-\infty}^{\infty}\,dx\,e^{-V\left( x\right)}v_i(x)v_j(x)=\delta_{ij},
\label{3.9.1}
\end{equation}
which implies that the polynomials $\{P_i (x)\}$ in (\ref{3.8}) are the set of
polynomials orthogonal on the real line with respect to the weight
$e^{-V\left(x\right)}$. In particular, for $V(x)=U(x)$, we have $\Psi_V(x)=
\Psi_U(x)\equiv\Psi_0(x)$ and $\{v_i(x)\}$ are identical to the orthonormal
functions $\{u_i(x)\}$ defined in (\ref{6}).

Consider the expectation value of the Hamiltonian ${\cal H}$ in (\ref{3.1})
in the normalizable state of (\ref{3.6}):
\begin{equation}
{\cal E}_0(V) = {1\over {\cal Z}_V}\int d^{N^2}\Phi\, e^{-\frac{1}{2}{\rm Tr}V}
{\cal H}e^{-\frac{1}{2}{\rm Tr}V},
\label{3.10}
\end{equation}
where
\begin{equation}
{\cal Z}_V =\int d^{N^2}\Phi\, e^{-{\rm Tr}V}\, =\,\langle \Psi_V|\Psi_V
\rangle<\infty.
\label{3.11}
\end{equation}
Clearly,
\begin{equation}
{\cal E}_0(V) \geq 0
\label{3.12}
\end{equation}
for any normalizable $\Psi_V$.

Since we have
\begin{equation}
{\cal E}_0(U) = 0,~~~ {\cal Z}_U < \infty,\nonumber
\end{equation}
we expect $\Psi_U\equiv\Psi_0$ to be an absolute quadratic minimum for
${\cal E}_0(V)$ in the space of functions of the form given by (\ref{3.6}) and
(\ref{3.11}). Indeed, for $V$ infinitesimally different from $U$,
\begin{equation}
V = U + \delta U,
\label{3.13}
\end{equation}
we find that
\begin{equation}
{\cal E}_0(U+\delta U)=\langle \Psi_U|\frac{1}{8} {\rm Tr}(\delta U^\prime)^2 |
\Psi_U \rangle/ {\cal Z}_U + {\cal O}(\delta U)^3 \geq 0.
\label{3.14}
\end{equation}
Expressing (\ref{3.14}) in matrix polar coordinates and using the Slater
determinant representation (\ref{3.9}) for $\Psi_V$ (and for $\Psi_U$),
we find, after some trivial manipulations, that
\begin{eqnarray}
&&{\cal E}_0(U + \delta U)=\sum^{N-1}_{i=0}\int dx
v_i^\ast(x)\left[-\frac{1}{2}
\frac{d^2}{dx^2}-\frac{1}{4}U^{\prime\prime}(x)+\frac{1}{8}U^{\prime^2}(x)+
\mu_i^{(N)}\right] v_i(x)\nonumber\\
&&\quad - \frac{1}{4} \Sigma_{i,j}^\prime \int dx dy\, v^\ast_i(x)v^\ast_j(y)
\frac{U^\prime(x)-U^\prime(y)}{x-y} [v_i(x)v_j(y)-v_j(x)v_i(y)]
-\sum^{N-1}_{i=0} \mu_i^{(N)}\nonumber\\
&&\quad\quad\quad =\frac{1}{8}\sum^{N-1}_{i=0}\int dx
|u_i(x)|^2 (\delta U^\prime(x))^2 + {\cal O}(\delta U)^3.
\label{3.15}
\end{eqnarray}
Here, the $\{\mu_i^{(N)}\}_{i=0}^{N-1}$ are a set of $N$ Lagrange multipliers
(also known as chemical potentials) that will enforce the unit normalization
condition on the $\{v_i\}$ in the variational calculation.\footnote{It is
unnecessary to include in (\ref{3.15}) Lagrange multipliers to enforce
{\sl all\/} orthonormality conditions (\ref{3.9.1}) because the $\{v_i\}$ which
result from the variational calculation turn out {\sl a posteriori\/} to be
orthonormal\cite {BJ}.}

Taking the variational derivative of (\ref{3.15}) with respect to
$v^\ast_i(x)$,
at $V=U$, we find that
\begin{eqnarray}
&&\delta {\cal E}_0(U + \delta V)/\delta v_i^\ast(x)|_{\delta U = 0}
= \{ [ - \frac{1}{2} \partial^2_x - \frac{1}{4}
U^{\prime\prime} + \frac{1}{8} U^{\prime^2}+\mu_i^{(N)}]v_i(x)\nonumber\\
&&\quad\quad -\frac{1}{2} \Sigma_j^\prime \int dy v_j^\ast(y) \frac{U^\prime(x)
-U^\prime(y)}{x-y}[ v_i(x)v_j(y)-v_j(x)v_i(y) ]\}|_{v_i=u_i}\nonumber\\
&&\quad\quad\quad\quad=\frac{1}{4}\sum^N_{i=1}\int dy |u_i(y)|^2
\delta U^\prime(y) \frac{\delta U^\prime(y)}{\delta
v_i^\ast(x)}|_{\delta U = 0} \equiv 0.
\label{3.16}
\end{eqnarray}
This expression must vanish because
\begin{eqnarray}
K_i(x,y)=\frac{\delta U^\prime(x)}{\delta v_i(y)}
\label{3.17}
\end{eqnarray}
must be a regular kernel in function space for $V$ infinitesimally close to $U$
because $V(x)$ defines the $v_i(x)$ uniquely, and vice-versa
[$V(x)=-2\log\left(
\frac{v_0(x)}{{\cal P}_0}\right)$]. At the point $V(x)=U(x)$ in function space
the set of functions $\{v_i\left(x\right)\}$ coincides with the set of
functions
$\{u_i\left(x\right)\}$ defined in (\ref{6}), where the $\{{\cal P}_i\left(x
\right)\}$ are the set of polynomials orthonormal with respect to the weight
$e^{-U\left( x\right)}$ on the real line. Moreover, uniqueness of the ground
state of the quantum-mechanical system defined by (\ref{3.1}) implies that
(\ref{6}) is the {\sl unique\/} solution to (\ref{1}). This completes our proof
of the assertion made in Eqs.~(\ref{1}) and (\ref{6}).

We conclude this section by proving that the first $\nu$ orthonormal
polynomials, $\cp_0$, $\cp_1$, $\cp_2$, $\ldots$, $\cp_\nu$, associated with
the
weight $w^2(x)=e^{-U(x)}$ satisfy the useful sum-rule
\begin{eqnarray}
&&\sum_{m,n=0}^{\nu-1}\int_{-\infty}^{\infty}dx\;w^2(x)\int_{-\infty}^{\infty}dy
\;w^2(y){U'(x)-U'(y)\over x-y}\cp_n(x)\cp_m(y)\left[\cp_n(x)\cp_m(y)-\cp_m(x)
\cp_n(y)\right]\nonumber\\
&&\quad\quad=2\sum_{n=0}^{\nu-1}\int_{-\infty}^{\infty}dx\;w^2(x)[\cp_n'(x)]^2.
\label{3.18}
\end{eqnarray}
To prove (\ref{3.18}) we let $\Phi$ be a $\nu\times\nu$ Hermitian matrix and
define
\begin{equation}
S_{\nu}={\int d^{\nu^2}\Phi e^{-{\rm Tr}U(\Phi)}{\rm Tr}\left[\left({\partial
\over\partial\Phi}\right) U'(\Phi)\right]\over\int d^{\nu^2}\Phi e^{-{\rm Tr}
U(\Phi)}}
\label{3.19}
\end{equation}

Integrating by parts over $\Phi$ we obtain straight forwardly
\begin{equation}
S_{\nu}={\int d^{\nu^2}\Phi e^{-{\rm Tr} U(\Phi)} {\rm Tr}\left[U'^2(\Phi)
\right]\over\int d^{\nu^2}\Phi e^{-{\rm Tr}U(\Phi)}}.
\label{3.20}
\end{equation}
In terms of the Slater determinant (\ref{3.9}) (with $V=U$) the latter equation
becomes\cite{BJ}
\begin{equation}
S_{\nu}=\sum_{n=0}^{\nu-1}\int_{-\infty}^{\infty}dx\;w^2(x)\cp_n^2(x)U'^2(x).
\label{3.21}
\end{equation}

On the other hand, using the identity
\begin{equation}
{\rm Tr}\left[\left({\partial\over\partial\Phi}\right)U'(\Phi)\right]=\sum_{m,n
=1}^{\nu}{U'(x_m)-U'(x_n)\over x_m-x_n}=\sum_{m=1}^{\nu}U^{\prime\prime}(x_m)
+\sum_{m\neq n=1}{U'(x_m)-U'(x_n)\over x_m-x_n}
\label{3.22}
\end{equation}
and the same Slater determinant as above, we obtain\cite{BJ} an alternative
expression for $S_{\nu}$ as
\begin{eqnarray}
&&S_{\nu}=
\nonumber\\
&&\sum_{m,n=0}^{\nu-1}\int_{-\infty}^{\infty}dx w^2(x)\int_{-\infty}^{\infty}dy
 w^2(y) {U'(x)-U'(y)\over x-y} \cp_n(x)\cp_m(y)\left[\cp_n(x)\cp_m(y)-\cp_m(x)
\cp_n(y)\right]\nonumber\\
&&\quad\quad  + \sum_{n=0}^{\nu-1}\int_{-\infty}^{\infty}dx w^2(x) \cp_n^2(x)
U^{\prime\prime}(x).
\label{3.23}
\end{eqnarray}
Equating (\ref{3.21}) and (\ref{3.23}), using the identity $e^{-U(x)}[U'^2(x)-
U^{\prime\prime}(x)]={d^2\over dx^2} e^{-U(x)}$ and the orthonormality
condition (\ref{5}), we obtain (\ref{3.18}) as required.
\pagebreak

\section{Eigenvalues of the Integro-Differential Equation}

In the previous section we have proved that the set of functions in (\ref{6})
is the unique solution of the system of integro-differential equations in
(\ref{1}). The eigenvalues $\{\mu_i^{\left(N\right)}\}$ of (\ref{1}) played
absolutely no role in that proof because we already knew the exact form
(\ref{3.2}) of the Fermi gas ground state and used it to show that (\ref{1})
and
(\ref{6}) hold. However, these eigenvalues are an indispensable ingredient of
the system (\ref{1}) because they are Lagrange multipliers that enforce the
orthonormality conditions (\ref{3.9.1}). They therefore encode important
information about the orthonormal polynomials (\ref{6}) in much the same way
such information is encoded by the set of moments of the weight
$e^{-U\left(x\right)}$ or by the set of coefficients of the generic three-term
recursion relations among the polynomials ${\cal P}_i$
\cite{Szego,Bateman,Mehta,BIZ}.\footnote{Note, however, that the latter two
sets
of coefficients involve {\sl all\/} polynomials, while the eigenvalues
$\{\mu_i^{\left(N\right)}\}$ do depend on the number N of polynomials used.}
Since the polynomials ${\cal P}_i$ are known once the weight $e^{-U(x)}$ is
given, we can use (\ref{1}) to determine the eigenvalues $\{\mu_i^{(N)}\}$.

Multiplying (\ref{1}) by $u_i^*(x)$ from (\ref{6}) and integrating over $x$
we find, using (\ref{2}), that the eigenvalues are given by
\begin{equation}
\mu_n^{(N)}=-{1\over 2}\int_{-\infty}^{\infty}dx\;w^2(x)\left({\cal P}_n^{
\prime}\right)^2+{1\over 2}\sum_{m=0}^{N-1} K_{nm}\quad (0\leq n \leq N-1).
\label{7.1}
\end{equation}
Here, as before, $w^2(x)=e^{-U(x)}$ and
\begin{eqnarray}
K_{nm}&=&K_{mn}\nonumber\\
&=&\int_{-\infty}^{\infty}dx\;w^2(x)\int_{-\infty}^{\infty}dy\; w^2(y)
{U'\left(x\right)-U'\left(y\right)\over x-y }\nonumber\\
&&\quad\quad\times\cp_n\left( x\right)\cp_m\left(y\right)\left[\cp_n\left(x
\right)\cp_m\left(y\right)-\cp_n\left(y\right)\cp_m\left(x\right)\right]
\nonumber\\
&\equiv &\int_{-\infty}^{\infty}dx\int_{-\infty}^{\infty}dy\;{\cal
K}_{nm}(x,y),
\label{7.2}
\end{eqnarray}
where we define
\begin{equation}
{\cal K}_{nm}(x,y) \equiv w^2(x) w^2(y)\left(\partial_x-\partial_y\right)
{\cp_n\left(x\right)\cp_m\left(y\right)\left[ \cp_n\left(x\right)\cp_m
\left(y\right) - \cp_n\left(y\right)\cp_m\left(x\right)\right]\over x-y }.
\label{7.2.1}
\end{equation}

It is obvious from (\ref{7.2}) that the symmetric matrix $K_{nm}$ has vanishing
diagonal elements
\begin{equation}
K_{nn}=0.
\label{7.3}
\end{equation}

The sum-rule (\ref{3.18}) we have found in the previous section now becomes
\begin{equation}
\sum_{n,\;m = 0}^{N-1} K_{nm} = 2\sum_{n=0}^{N-1}\int_{-\infty}^{\infty}
dx\; w^2(x)\left(\cp_n^{\prime}\right)^2.
\label{7.4}
\end{equation}
Using (\ref{7.2})-(\ref{7.4}) and the symmetry of the matrix $K$ we obtain
after some algebra yet another sum rule
\begin{equation}
\sum_{m=0}^{N-1}K_{nm}=\sum_{k=n}^{N-1}\int_{-\infty}^{\infty}dx\;w^2(x)
\left(\cp_k^{\prime}\right)^2-\sum_{k=n+1}^{N-1}\sum_{l=0}^{N-1} K_{kl}
+{1\over 2}\sum_{k,l=n}^{N-1}K_{kl}\quad (n\leq N-2).
\label{7.5}
\end{equation}
 From (\ref{7.1}) and (\ref{7.5}) we infer the following recursion relation
satisfied by the eigenvalues
\begin{eqnarray*}
\mu_n^{(N)} = -\sum_{k=n+1}^{N-1} \mu_k^{(N)} + {1\over 4}\sum_{k,l=n}^{N-1}
K_{kl}\quad (n\leq N-2).
\end{eqnarray*}
The symmetry of the matrix $K$ allows us to restrict the double sum on
the right side of the last equation to the upper triangular part of $K$;
the recursion relation then becomes
\begin{equation}
\mu_n^{(N)} = -\sum_{k=n+1}^{N-1} \mu_k^{(N)} + {1\over 2}\sum_{k=n}^{N-1}
\sum_{p=1}^{N-k-1} K_{k,k+p}\quad (n\leq N-2).
\label{7.6}
\end{equation}
This recursion relation may be used to calculate $\mu_n^{(N)}$ for descending
values of $n$ holding $N$ fixed. Similarly, (\ref{7.1}) implies that along the
$N$ index direction there is a recursion relation of the form
\begin{equation}
\mu_n^{(N+1)}=\mu_n^{(N)}+{1\over 2}K_{n,N}\quad(0\leq n\leq N-1).
\label{7.7}
\end{equation}
The equations (\ref{7.7}) and (\ref{7.6}) are not independent and thus they
cannot be combined to give a single self-contained recursion relation in terms
of the eigenvalues $\mu_n^{(N)}$ alone. Nevertheless, we {\sl can} calculate
$\mu_n^{(N)}$ iteratively using the above formulas.

The initial condition for both (\ref{7.6}) and (\ref{7.7}), namely the value
of $\mu_{N-1}^{(N)}$, must be calculated separately. We show below that
\begin{equation}
\mu_{N-1}^{(N)}\equiv 0
\label{7.8}
\end{equation}
for {\sl any\/} even weight $w^2(x)=e^{-U(x)}$.

To prove this identity we note from (\ref{7.1}) that
\begin{equation}
\mu_{N-1}^{(N)}=-{1\over 2}\int_{-\infty}^{\infty}dx\;w^2(x)\left(
{\cal P}_{N-1}^{\prime}\right)^2+{1\over 2}\sum_{m=0}^{N-1} K_{N-1,m}.
\label{7.8.1}
\end{equation}

Rewriting the sum in the last equation as
\begin{equation}
\sum_{m=0}^{N-1}K_{N-1,m}=\left(\sum_{n,m=0}^{N-1}-\sum_{n,m=0}^{N-2}\right)
K_{nm}-\sum_{m=0}^{N-1}K_{m,N-1},
\label{7.8.2}
\end{equation}
and using the sum-rule (\ref{7.4}) and the symmetry of the matrix $K$,
we find that
\begin{equation}
\sum_{m=0}^{N-1} K_{N-1,m}=2\int_{-\infty}^{\infty}dx\;w^2(x)\left(
{\cal P}_{N-1}^{\prime}\right)^2-\sum_{m=0}^{N-1}K_{N-1,m}.
\nonumber
\end{equation}
Therefore,
\begin{equation}
\sum_{m=0}^{N-1}K_{N-1,m}=\int_{-\infty}^{\infty}dx\;w^2(x)\left(
{\cal P}_{N-1}^{\prime}\right)^2
\label{7.8.3}
\end{equation}
which upon substitution into (\ref{7.8.1}), yields the result (\ref{7.8}).

We now give explicit general formulas for the first few eigenvalues
$\mu_n^{(N)}$. Our procedure is to use (\ref{7.2}) and (\ref{7.2.1}) to obtain
a
sequence of formulas for $K_{nm}$ and thus for the eigenvalues. These formulas
are obtained from the polynomial recursion relation
\cite{Mehta,BIZ,Szego,Bateman}
\begin{equation}
x\cp_n (x)=\sqrt{R_{n+1}}\;\cp_{n+1}(x)+\sqrt{R_n}\;\cp_{n-1}(x),
\label{7.9}
\end{equation}
where the recursion coefficients $R_n$ are uniquely defined by the weight
function $w^2(x)$. Note that this recursion relation generates normalized
polynomials $\cp_n(x)$:
\begin{eqnarray}
\cp_0(x)&=&1,\nonumber\\
\cp_1(x)&=&{x\over\sqrt{R_1}},\nonumber\\
\cp_2(x)&=&{x^2-R_1\over\sqrt{R_1R_2}},\nonumber\\
\cp_3(x)&=&{x^3-(R_1+R_2)x\over\sqrt{R_1R_2R_3}},\nonumber\\
\cp_4(x)&=&{x^4-(R_1+R_2+R_3)x^2+R_1R_3\over\sqrt{R_1R_2R_3R_4}},\nonumber\\
\cp_5(x)&=&{x^5-(R_1+R_2+R_3+R_4)x^3+(R_1R_3+R_1R_4+R_2R_4)x\over
\sqrt{R_1R_2R_3R_4R_5}},\nonumber\\
\cp_6(x)&=&{1\over\sqrt{R_1R_2R_3R_4R_5R_6}}[x^6-(R_1+R_2+R_3+R_4+R_5)x^4
\nonumber\\
&&\quad\quad +(R_1R_3+R_1R_4+R_1R_5+R_2R_4+R_2R_5+R_3R_5)x^2-R_1R_3R_5],
\nonumber\\
\cp_7(x)&=&{1\over\sqrt{R_1R_2R_3R_4R_5R_6}}[x^7-(R_1+R_2+R_3+R_4+R_5+R_6)x^5
+(R_1R_3\nonumber\\
&&\quad\quad
+R_1R_4+R_1R_5+R_1R_6+R_2R_4+R_2R_5+R_2R_6+R_3R_5+R_3R_6\nonumber\\
&&\quad\quad +R_4R_6)x^3 -(R_1R_3R_5+R_1R_3R_6+R_1R_4R_6+R_2R_4R_6)x].
\label{7.10}
\end{eqnarray}

Inserting the above recursion relation into (\ref{7.2.1}) repeatedly allows
us to cancel the quantity $x-y$ in the denominator.\footnote{For $|m-n|=1$ in
(\ref{7.2}),(\ref{7.2.1}) this iterative process yields the Darboux-Christoffel
formula. When $|m-n|>1$ one obtains generalizations thereof
\cite{Szego,Bateman}.} The orthonormality relation
\begin{equation}
\int_{-\infty}^{\infty} dx\;w^2(x)\cp_n (x)\cp_m(x)= \delta_{nm}
\label{7.11}
\end{equation}
then simplifies the resulting expression dramatically. To present our final
results we define a set of polynomials $\cq_n^{(N)}(x)$ that are conjugate to
the polynomials $\cp_n(x)$:
\begin{eqnarray}
\cq_0^{(N)}(x)&=&1,\nonumber\\
\cq_1^{(N)}(x)&=&{x\over\sqrt{R_{N-1}}},\nonumber\\
\cq_2^{(N)}(x)&=&{x^2-R_{N-1}\over\sqrt{R_{N-1}R_{N-2}}},\nonumber\\
\cq_3^{(N)}(x)&=&{x^3-(R_{N-1}+R_{N-2})x\over\sqrt{R_{N-1}R_{N-2}R_{N-3}}},
\nonumber\\
\cq_4^{(N)}(x)&=&{x^4-(R_{N-1}+R_{N-2}+R_{N-3})x^2+R_{N-1}R_{N-3}\over
\sqrt{R_{N-1} R_{N-2}R_{N-3}R_{N-4}}},\nonumber\\
\cq_5^{(N)}(x)&=&{1\over\sqrt{R_{N-1}R_{N-2}R_{N-3}R_{N-4}R_{N-5}}}[x^5-(R_{N-1}
+R_{N-2}+R_{N-3}+R_{N-4})x^3\nonumber\\
&&\quad\quad +(R_{N-1}R_{N-3}+R_{N-1}R_{N-4}+R_{N-2}R_{N-4})x],\nonumber\\
\cq_6^{(N)}(x)&=&{1\over\sqrt{R_{N-1}R_{N-2}R_{N-3}R_{N-4}R_{N-5}R_{N-6}}}[
x^6-(R_{N-1}+R_{N-2}+R_{N-3}+R_{N-4}\nonumber\\
&&\quad\quad +R_{N-5})x^4+(R_{N-1}R_{N-3}+R_{N-1}R_{N-4}+R_{N-1}R_{N-5}
+R_{N-2}R_{N-4}\nonumber\\
&&\quad\quad +R_{N-2}R_{N-5}+R_{N-3}R_{N-5})x^2-R_{N-1}R_{N-3}R_{N-5}],
\nonumber\\
\cq_7^{(N)}(x)&=&{1\over\sqrt{R_{N-1}R_{N-2}R_{N-3}R_{N-4}R_{N-5}R_{N-6}}}[
x^7-(R_{N-1}+R_{N-2}+R_{N-3}+R_{N-4}\nonumber\\
&&\quad\quad +R_{N-5}+R_{N-6})x^5+(R_{N-1}R_{N-3}+R_{N-1}R_{N-4}+R_{N-1}R_{N-5}
\nonumber\\
&&\quad\quad +R_{N-1}R_{N-6}+R_{N-2}R_{N-4}+R_{N-2}R_{N-5}+R_{N-2}R_{N-6}
+R_{N-3}R_{N-5}\nonumber\\
&&\quad\quad +R_{N-3}R_{N-6}+R_{N-4}R_{N-6})x^3-(R_{N-1}R_{N-3}R_{N-5}
\nonumber\\
&&\quad\quad +R_{N-1}R_{N-3}R_{N-6}+R_{N-1}R_{N-4}R_{N-6}+R_{N-2}R_{N-4}
R_{N-6})x],
\label{7.12}
\end{eqnarray}
where we have replaced $R_n$ in (\ref{7.10}) by $R_{N-n}$. For each value of
$N$ this set of polynomials is orthonormal as can be seen from the fact that
they satisfy a three-term recursion relation whose recursion coefficients
are $R_{N-n}$.

We then obtain the formula
\begin{equation}
K_{n,n-k}={1\over\sqrt{R_n}}\int_{-\infty}^{\infty}dx\;w^2(x)\cp_n'(x)
\cq_{k-1}^{(n)}(x)\cp_{n-k}(x).
\label{7.13}
\end{equation}

Once we have determined the matrix $K_{nm}$ we can then use (\ref{7.7}) to
construct a formula for the eigenvalues $\mu_n^{(N)}$:
\begin{equation}
\mu_n^{(N)}={1\over 2}\sum_{j=1}^{N-n-1} K_{n,N-j}\quad (n\leq N-1).
\label{7.14}
\end{equation}

We have been able to find closed-form expressions for some of the $K_{nm}$:
\begin{eqnarray}
K_{n,n-1}&=&{n\over R_n}\quad (n\geq 1),\nonumber\\
K_{n,n-2}&=&{2\over R_n R_{n-1}}\sum_{k=1}^{n-1}R_k\quad (n\geq 2),\nonumber\\
K_{n,n-3}&=&{1\over R_n R_{n-1}R_{n-2}}\sum_{k=1}^{n-2}[R_{n-1}^2+2R_k^2+4R_k
R_{k+1}-4R_{n-1}R_k]\quad (n\geq 3),\nonumber\\
K_{n,n-4}&=&{2\over R_n R_{n-1}R_{n-2}R_{n-3}}\sum_{k=1}^{n-3}[(R_{n-1}+
R_{n-2})^2 R_k-2(R_{n-1}+R_{n-2})(R_k^2+2R_kR_{k+1})\nonumber\\
&&\quad\quad +3R_kR_{k+1}R_{k+2}+R_k^3+3R_k^2R_{k+1}+3R_kR_{k+1}^2]\quad (n
\geq 4),
\label{7.15}
\end{eqnarray}
and so on. These formulas become increasingly complicated as one moves further
away from the diagonal of the matrix $K$.

 From (\ref{7.14}) and (\ref{7.15}) we obtain the following formulas for the
first few eigenvalues:
\begin{eqnarray}
\mu_{N-1}^{(N)}&=&0\quad (N\geq 1),\nonumber\\
\mu_{N-2}^{(N)}&=&{N-1\over 2 R_{N-1}}\quad (N\geq 2),\nonumber\\
\mu_{N-3}^{(N)}&=&{1\over 2 R_{N-1}R_{N-2}}[2(R_1+\dots+R_{N-2})+(N-2)R_{N-1}]
\quad (N\geq 3),
\label{7.16}
\end{eqnarray}
and so on. Again, these formulas become more complicated as the difference
between the two indices of $\mu$ grows.
\pagebreak

\section{Eigenvalues of the Hahn-Meixner Polynomials}

In the previous section our discussion was completely general; it applies to
the
case of an even weight function $w^2(x)$ on the whole real line. To illustrate
the results of the previous section we now focus on a special class of
polynomials known as the Hahn-Meixner polynomials. In the context of the
discussion above, these polynomials are single-particle states from which one
constructs the ground state of a one-dimensional Fermi gas.

The Hahn-Meixner polynomials \cite{Hahn,Askey1,Askey2,AS,BMP1,Kniga} are a
four-parameter family of polynomials $\cp_n(x)$ orthogonal on the whole real
line ($-\infty<x<\infty$). These polynomials are expressible in terms of a
generalized hypergeometric function:
\begin{equation}
\cp_n(x)=N_n i^n \,{}_3 F_2 (-n,n+a+b+c+d-1,a-ix;a+c,a+d;1),
\label{h1}
\end{equation}
where $a$, $b$, $c$, and $d$ are parameters that may take on any complex values
and the normalization $N_n$ is given by
\begin{equation}
N_n = {[(2n+a+b+c+d-1)\Gamma(n+a+b+c+d-1)\Gamma(n+a+c)\Gamma(n+a+d)]^{1/2}
\over \Gamma(a+c)\Gamma(a+d) n!\,[\Gamma(n+b+c)\Gamma(n+b+d)]^{1/2}}.
\label{h2}
\end{equation}
These polynomials are orthonormal on the real line
\begin{equation}
\int_{-\infty}^{\infty}dx\, w^2(x) {\cal P}_m(x){\cal P}_n(x)=\delta_{mn},
\label{h3}
\end{equation}
where the weight $w^2(x)$ is given by
\begin{equation}
w^2(x) = {1\over 2\pi} \Gamma(a+ix)\Gamma(b+ix)\Gamma(c-ix)\Gamma(d-ix).
\label{h4}
\end{equation}

Hahn-Meixner polynomials appear often in mathematical physics; for example,
they
play a major role in the representation theory of the Lorentz and rotation
groups \cite{Kniga,Bhn}. Recently, it has been shown that a special case of the
Hahn-Meixner polynomials plays a crucial role in discrete-time quantum
mechanics \cite{BMP2}, the solution of operator differential equations
\cite{BD}, the operator ordering problem  in quantum mechanics \cite{BMP2}, and
in the theory of Weyl-ordered (symmetric) operators of the Heisenberg algebra
\cite{BMP1}. Here we restrict our attention to the special case of Hahn-Meixner
polynomials that appear in the latter applications. These particular
polynomials
are characterized by the parameter values
\begin{equation}
a=c={1\over 4}\quad {\rm and}\quad b=d={3\over 4}
\label{h5}
\end{equation}
in (\ref{h1}).

This special case of Hahn-Meixner polynomials may be obtained in a natural way
from the Heisenberg algebra as follows. We introduce a complete operator basis
having a simple commutator algebra. This Hermitian basis $T_{m,n}$ is the sum
over all possible orderings of $m$ operators $p$ and $n$ operators $q$. For
example,
\begin{eqnarray}
T_{1,1}&=&pq+qp,\nonumber\\
T_{0,1}&=&q,\nonumber\\
T_{2,1}&=&p^2q+pqp+qp^2,\nonumber\\
T_{2,2}&=&(p^2 q^2 +q^2 p^2 + p q^2 p+qp^2 q+qpqp+pqpq).
\label{h6}
\end{eqnarray}
These symmetrized operator products $T_{m,n}$ can be rewritten as
Weyl-ordered sums in the operators $p$ and $q$ with the terms in the sums
weighted by binomial coefficients \cite{L,Mish}:
\begin{equation}
T_{m,n}={(m+n)!\over 2^n m!\,n!}\sum_{k=0}^n\left ({n\atop k}\right )
q^k p^m q^{(n-k)}.
\label{h7}
\end{equation}

These basis element operators $T_{m,n}$ have particularly simple commutation
properties:
\begin{eqnarray}
[q,T_{m,n}] &=& i(m+n) T_{m-1,n},\cr
[p,T_{m,n}] &=& -i(m+n) T_{m,n-1},\cr
\{q,T_{m,n}\}_+ &=&{2(n+1)\over m+n+1} T_{m,n+1},\cr
\{p,T_{m,n}\}_+ &=&{2(m+1)\over m+n+1} T_{m+1,n}.
\label{h8}
\end{eqnarray}
Thus, commuting with $q$ and $p$ has the effect of a lowering operator and
anticommuting with $q$ and $p$ has the effect of a raising operator in the
appropriate index. From the four basic algebraic relations (\ref{h8}) one can
deduce a general formula for the commutation and anticommutation relations
between $T_{m,n}$ and $T_{r,s}$\cite{BD2}.

The connection between the totally symmetrized operators $T_{m,n}$ and the
special case of the Hahn-Meixner polynomials in (\ref{h1}) and (\ref{h5}) is
\begin{eqnarray}
T_{n,n} &=& {1\over (2n-1)!!} \cp_n(T_{1,1}),\cr
T_{n,n+k} &=& {(2n+k)!\over(n+k)!2^{n+1}}\{ q^k, \cp_n(T_{1,1}) \}_+ .
\label{e9}
\end{eqnarray}

We list below the properties of the special Hahn-Meixner polynomials:

\noindent
 ~~(i)~~The $\cp_n(x)$ satisfy the recurrence relation
\begin{eqnarray}
n\cp_n(x) = x \cp_{n-1}(x) - (n-1)\cp_{n-2}(x).
\label{e9.1}
\end{eqnarray}
By comparing this special case with the general recursion relation in
(\ref{7.9}) we identify
\begin{eqnarray}
R_n = n^2.
\label{e9.2}
\end{eqnarray}

\noindent
 ~~(ii)~~The $\cp_n(x)$ are orthonormal on the interval $(-\infty,\infty)$
with respect to the weight function
\begin{equation}
e^{-U(x)}=w^2(x)={1\over 2 \cosh(\pi x/2)}.
\label{weight}
\end{equation}
The moments of this weight function are Euler numbers:
\begin{equation}
\int_{-\infty}^{\infty}dx\;w^2(x)x^{2n}=|E_{2n}|,
\label{euler}
\end{equation}
where $E_0=1$, $E_2=-1$, $E_4=5$, $E_6=-61$, $E_8=1385$, $E_{10}=-50521$, and
so
on.

\noindent
 ~~(iii)~~The $\cp_n(x)$ have a simple generating function
\begin{eqnarray}
\sum_{n=0}^{\infty} t^n\cp_n(x)={\exp (x\,{\rm arctan}\, t)\over\sqrt{1+t^2}}.
\nonumber
\end{eqnarray}

\noindent
 ~~(iv)~~The first few $\cp_n(x)$ are
\begin{eqnarray}
\cp_0(x) &=& 1,\nonumber\\
\cp_1(x) &=& x,\nonumber\\
\cp_2(x) &=& {1\over 2}\left (x^2-{1\over 2}\right ),\nonumber\\
\cp_3(x) &=& {1\over 6}\left (x^3-2x\right ),\nonumber\\
\cp_4(x) &=& {1\over 24}\left (x^4-5x^2+{3\over 2}\right ),\nonumber\\
\cp_5(x) &=& {1\over 5!}\left (x^5-10x^3+{23\over 2}x\right ),\nonumber\\
\cp_6(x) &=& {1\over 6!}\left (x^6-{35\over 2}x^4+49 x^2-{45\over 4}
\right ),\nonumber\\
\cp_7(x) &=& {1\over 7!}\left (x^7-28x^5+154x^3-132 x\right ),\nonumber\\
\cp_8(x) &=& {1\over 8!}\left (x^8-42x^6+399x^4-818x^2+{315\over 2}\right ).
\nonumber
\end{eqnarray}

Using these polynomials in (\ref{7.13}) we obtain the matrix $K_{mn}$, whose
first few entries are listed in Table 1.
\bigskip

{\baselineskip=24pt
\setbox\strutbox=\hbox{\vrule height11pt depth5pt width0pt}%
{\def\tablerule{\noalign{\hrule}}
\def\hwedge{\hbox to4pt{\hfill}}
\hoffset=.25in
\font\ninerm=cmr9
\overfullrule=0pt
\hbox to \hsize{\hfill}
\vbox{
\catcode`?=\active
\def?{\kern\digitwidth}
\openup1\jot \tabskip=0pt \offinterlineskip
\halign to \hsize{\strut#&\vrule#\tabskip=3pt plus1pt&#\hfil&\vrule#&
\hfil#\hfil&\hfil#\hfil&\hfil#\hfil&\hfil#\hfil&\hfil#\hfil&\hfil#\hfil&
\hfil#\hfil&\hfil#\hfil&\vrule#\tabskip=0pt\cr
\tablerule
\omit&height 3pt&\omit&&\omit&\omit&\omit&\omit&\omit&\omit&\omit&\omit&\cr
\omit&height 3pt&\omit&&\omit&\omit&\omit&\omit&\omit&\omit&\omit&\omit&\cr
&&\hfill $$&&m=0&m=1&m=2&m=3&m=4&m=5&m=6&m=7&\cr
\omit&height 3pt&\omit&&\omit&\omit&\omit&\omit&\omit&\omit&\omit&\omit&\cr
\tablerule
\omit&height 3pt&\omit&&\omit&\omit&\omit&\omit&\omit&\omit&\omit&\omit&\cr
&&$n=0$&&{\ninerm 0}&{\ninerm 1}&$1\over 2$&$1\over 2$&$13\over 36$&$7
\over 20$&$17\over 60$&$1727\over 6300$&\cr
\omit&height 3pt&\omit&&\omit&\omit&\omit&\omit&\omit&\omit&\omit&\omit&\cr
&&$n=1$&&{\ninerm 1}&{\ninerm 0}&$1\over 2$&$5\over 18$&$11\over 36$&$41
\over 180$&$209\over 900$&$1201\over 6300$&\cr
\omit&height 3pt&\omit&&\omit&\omit&\omit&\omit&\omit&\omit&\omit&\omit&\cr
&&$n=2$&&$1\over 2$&$1\over 2$&{\ninerm 0}&$1\over 3$&$7\over 36$&$67
\over 300$&$17\over 100$&$1121\over 6300$&\cr
\omit&height 3pt&\omit&&\omit&\omit&\omit&\omit&\omit&\omit&\omit&\omit&\cr
&&$n=3$&&$1\over 2$&$5\over 18$&$1\over 3$&{\ninerm 0}&$1\over 4$&$3
\over 20$&$53\over 300$&$6017\over 44100$&\cr
\omit&height 3pt&\omit&&\omit&\omit&\omit&\omit&\omit&\omit&\omit&\omit&\cr
&&$n=4$&&$13\over 36$&$11\over 36$&$7\over 36$&$1\over 4$&{\ninerm 0}&$1
\over 5$&$11\over 90$&$461\over 3150$&\cr
\omit&height 3pt&\omit&&\omit&\omit&\omit&\omit&\omit&\omit&\omit&\omit&\cr
&&$n=5$&&$7\over 20$&$41\over 180$&$67\over 300$&$3\over 20$&$1\over 5$&
{\ninerm 0}&$1\over 6$&$13\over 126$&\cr
\omit&height 3pt&\omit&&\omit&\omit&\omit&\omit&\omit&\omit&\omit&\omit&\cr
&&$n=6$&&$17\over 60$&$209\over 900$&$17\over100$&$53\over 300$&$11\over
90$&$1\over 6$&{\ninerm 0}&$1\over 7$&\cr
\omit&height 3pt&\omit&&\omit&\omit&\omit&\omit&\omit&\omit&\omit&\omit&\cr
&&$n=7$&&$1727\over 6300$&$1201\over 6300$&$1121\over 6300$&$6017
\over 44100$&$461 \over 3150$&$13\over 126$&$1\over 7$&{\ninerm 0}&\cr
\omit&height 3pt&\omit&&\omit&\omit&\omit&\omit&\omit&\omit&\omit&\omit&\cr
\omit&height 3pt&\omit&&\omit&\omit&\omit&\omit&\omit&\omit&\omit&\omit&\cr
\tablerule
}}}

\medskip
{\baselineskip=12pt
\noindent
{\bf Table~1}~~The first few entries of the symmetric matrix $K_{mn}$
associated with the Hahn-Meixner polynomials. The entries in this matrix
were computed from (\ref{7.13}).
\par}}

Applying (\ref{7.14}) to the entries in Table 1 we can compute the eigenvalues
$\mu_n^{(N)}$ associated with the Hahn-Meixner polynomials. The first few
eigenvalues are listed in Table 2.

\bigskip
{\setbox\strutbox=\hbox{\vrule height11pt depth5pt width0pt}%
{\def\tablerule{\noalign{\hrule}}
\def\hwedge{\hbox to4pt{\hfill}}
\hoffset=.25in
\font\ninerm=cmr9
\overfullrule=0pt
\hbox to \hsize{\hfill}
\vbox{
\catcode`?=\active
\def?{\kern\digitwidth}
\openup1\jot \tabskip=0pt \offinterlineskip
\halign to \hsize{\strut#&\vrule#\tabskip=3pt plus1pt&#\hfil&\vrule#&
\hfil#\hfil&\hfil#\hfil&\hfil#\hfil&\hfil#\hfil&\hfil#\hfil&\hfil#\hfil&
\hfil#\hfil&\hfil#\hfil&\hfil#\hfil&\hfil#\hfil&\hfil#\hfil&
\vrule#\tabskip=0pt\cr
\tablerule
\omit&height 3pt&\omit&&\omit&\omit&\omit&\omit&\omit&\omit&\omit&\omit&\omit&
\omit&\omit&\cr
\omit&height 3pt&\omit&&\omit&\omit&\omit&\omit&\omit&\omit&\omit&\omit&\omit&
\omit&\omit&\cr
&&\hfill $n$&&0&1&2&3&4&5&6&7&8&9&10&\cr
\omit&height 3pt&\omit&&\omit&\omit&\omit&\omit&\omit&\omit&\omit&\omit&\omit&
\omit&\omit&\cr
\tablerule
\omit&height 3pt&\omit&&\omit&\omit&\omit&\omit&\omit&\omit&\omit&\omit&\omit&
\omit&\omit&\cr
&&$N=1$&&{\ninerm 0}& & & & & & & & & & &\cr
\omit&height 3pt&\omit&&\omit&\omit&\omit&\omit&\omit&\omit&\omit&\omit&\omit&
\omit&\omit&\cr
&&$N=2$&&$1\over 2$&{\ninerm 0}& & & & & & & & & &\cr
\omit&height 3pt&\omit&&\omit&\omit&\omit&\omit&\omit&\omit&\omit&\omit&\omit&
\omit&\omit&\cr
&&$N=3$&&$3\over 4$&$1\over 4$&{\ninerm 0}& & & & & & & & &\cr
\omit&height 3pt&\omit&&\omit&\omit&\omit&\omit&\omit&\omit&\omit&\omit&\omit&
\omit&\omit&\cr
&&$N=4$&&{\ninerm 1}&$7\over 18$&$1\over 6$&{\ninerm 0}& & & & & & & &\cr
\omit&height 3pt&\omit&&\omit&\omit&\omit&\omit&\omit&\omit&\omit&\omit&\omit&
\omit&\omit&\cr
&&$N=5$&&$85\over 72$&$13\over 24$&$19\over 72$&$1\over 8$&{\ninerm 0}& & & &
 & & &\cr
\omit&height 3pt&\omit&&\omit&\omit&\omit&\omit&\omit&\omit&\omit&\omit&\omit&
\omit&\omit&\cr
&&$N=6$&&$61\over 45$&$59\over 90$&$169\over 450$&$1\over 5$&$1\over 10$&
{\ninerm 0}& & & & & &\cr
\omit&height 3pt&\omit&&\omit&\omit&\omit&\omit&\omit&\omit&\omit&\omit&\omit&
\omit&\omit&\cr
&&$N=7$&&$539\over 360$&$463\over 600$&$829\over1800$&$173\over 600$&$29\over
180$&$1\over 12$&{\ninerm 0}& & & & &\cr
\omit&height 3pt&\omit&&\omit&\omit&\omit&\omit&\omit&\omit&\omit&\omit&\omit&
\omit&\omit&\cr
&&$N=8$&&$286\over 175$&$2731\over 3150$&$577\over1050$&$3931\over 11025$&$41
\over 175$&$17\over 126$&$1\over 14$&{\ninerm 0}& & & &\cr
\omit&height 3pt&\omit&&\omit&\omit&\omit&\omit&\omit&\omit&\omit&\omit&\omit&
\omit&\omit&\cr
&&$N=9$&&$14713\over 8400$&$8083\over 8400$&$7331\over 11760$&$75703\over
176400$&$51403\over 176400$&$199\over 1008$&$13\over 112$&$1\over 16$&
{\ninerm 0}& & &\cr
\omit&height 3pt&\omit&&\omit&\omit&\omit&\omit&\omit&\omit&\omit&\omit&\omit&
\omit&\omit&\cr
&&$N=10$&&$3917\over 2100$&$46057\over 44100$&$489\over 700$&$21607\over
44100$&$20009\over 56700$&$145\over 588$&$43\over 252$&$11\over 108$&
$1\over 18$&{\ninerm 0}& &\cr
\omit&height 3pt&\omit&&\omit&\omit&\omit&\omit&\omit&\omit&\omit&\omit&\omit&
\omit&\omit&\cr
&&$N=11$&&$346753\over 176400$&$22067\over 19600$&$134737\over 176400$&$877307
\over 1587600$&$214189\over 529200$&$9257\over 317520$&$22633\over 105840$&$541
\over 3600$&$49\over 540$&$1\over 20$&{\ninerm 0}&\cr
\omit&height 3pt&\omit&&\omit&\omit&\omit&\omit&\omit&\omit&\omit&\omit&\omit&
\omit&\omit&\cr
\omit&height 3pt&\omit&&\omit&\omit&\omit&\omit&\omit&\omit&\omit&\omit&\omit&
\omit&\omit&\cr
\tablerule
}}}

\medskip
{\baselineskip=12pt
\noindent
{\bf Table~2}~~Table of the eigenvalues $\mu_n^{(N)}$ associated with the
Hahn-Meixner polynomials. These eigenvalues are obtained from the entries in
Table 1 using (\ref{7.14}).
\par}
\smallskip }

For the case of the Hahn-Meixner polynomials the general formulas in
(\ref{7.16}) become
\begin{eqnarray}
\mu_{N-1}^{(N+1)}&=&{1\over 2N},\nonumber\\
\mu_{N-2}^{(N+1)}&=&{5N-1 \over 6(N-1)N},\nonumber\\
\mu_{N-3}^{(N+1)}&=&{19N^2-28N+3\over 15(N-2)(N-1)N},\nonumber\\
\mu_{N-4}^{(N+1)}&=&{171N^3-633N^2+551N-45\over
105(N-3)(N-2)(N-1)N},\nonumber\\
\mu_{N-5}^{(N+1)}&=&{1289N^4-8980N^3+19351N^2-13052N+840\over
630(N-4)(N-3)(N-2)(N-1)N},\nonumber\\
\mu_{N-6}^{(N+1)}&=&{16757N^5-187721N^4+738847N^3-1201015N^2+689652N-37800\over
6930(N-5)(N-4)(N-3)(N-2)(N-1)N},\nonumber\\
\mu_{N-7}^{(N+1)}&=&[2(63705N^6-1048800N^5+6531490N^4-19071030N^3+25902989N^2
\nonumber\\
&&\quad\quad -13249854N+623700)]\nonumber\\
&&\quad\quad /[45045(N-6)(N-5)(N-4)(N-3)(N-2)(N-1)N],\nonumber\\
\mu_{N-8}^{(N+1)}&=&[2(72199N^7-1639133N^6+14772758N^5-67196100N^4+161413785N^3]\nonumber\\
&&\quad\quad -192719779N^2+90335010N-3783780)]\nonumber\\
&&\quad\quad /[45045(N-7)(N-6)(N-5)(N-4)(N-3)(N-2)(N-1)N],
\label{impressive}
\end{eqnarray}
and so on.

We have discovered that these formulas simplify considerably if they are
rewritten in the form of partial fractions:
\begin{eqnarray}
2\cdot 1!!\mu_{N-1}^{(N+1)}&=&{1\over N},\nonumber\\
2\cdot 3!! \mu_{N-2}^{(N+1)}&=&{1\over N}+{4\over N-1},\nonumber\\
2\cdot 5!!\mu_{N-3}^{(N+1)}&=&{3\over N}+{12\over N-1}+{23\over
N-2},\nonumber\\
2\cdot 7!!\mu_{N-4}^{(N+1)}&=&{15\over N}+{44\over N-1}+{107\over N-2}
+{176\over N-3},\nonumber\\
2\cdot 9!!\mu_{N-5}^{(N+1)}&=&{105\over N}+{276\over N-1}+{693\over N-2}
+{1104\over N-3}+{1689\over N-4},\nonumber\\
2\cdot 11!!\mu_{N-6}^{(N+1)}&=&{945\over N}+{2340\over N-1}+{4773\over N-2}
+{9360\over N-3}+{13329\over N-4}+{19524\over N-5},\nonumber\\
2\cdot 13!!\mu_{N-7}^{(N+1)}&=&{10395\over N}+{24780\over N-1}+{46767\over N-2}
+{94512\over N-3}+{138027\over N-4}+{185772\over N-5}+{264207\over N-6},
\nonumber\\
2\cdot 15!!\mu_{N-8}^{(N+1)}&=&{135135\over N}+{313740\over N-1}+{566235\over
N-2}+{978480\over N-3}+{1694655\over N-4}+{2264940\over N-5}\nonumber\\
&&\quad\quad +{2944395\over N-6}+{4098240\over N-7}.\nonumber\\
\label{nice}
\end{eqnarray}
Moreover, we have found that many of the coefficients can be expressed using
very simple formulas. For example, the leftmost coefficient is a double
factorial: $1=(-1)!!$, $1=1!!$, $3=3!!$, $15=5!!$, $105=7!!$, $945=9!!$,
$10395=11!!$, and $135135=13!!$. Also, the rightmost coefficient is the sum
of inverse odd integers:
\begin{eqnarray}
1 &=& 1!!\left ({1\over 1}\right ) ,\nonumber\\
4 &=& 3!!\left ({1\over 1}+{1\over 3}\right ) ,\nonumber\\
23 &=& 5!!\left ({1\over 1}+{1\over 3}+{1\over 5}\right ) ,\nonumber\\
176 &=&7!!\left({1\over 1}+{1\over 3}+{1\over 5}+{1\over 7}\right ),\nonumber\\
1689 &=& 9!!\left ({1\over 1}+{1\over 3}+{1\over 5}+{1\over 7}+{1\over 9}
\right ) ,\nonumber\\
19524 &=& 11!!\left ({1\over 1}+{1\over 3}+{1\over 5}+{1\over 7}+{1\over 9}
+{1\over 11}\right ) ,\nonumber\\
26247 &=& 13!!\left ( {1\over 1}+{1\over 3}+{1\over 5}+{1\over 7}+{1\over 9}
+{1\over 11}+{1\over 13}\right ) ,\nonumber\\
4098240 &=& 15!!\left ( {1\over 1}+{1\over 3}+{1\over 5}+{1\over 7}+{1\over 9}
+{1\over 11}+{1\over 13}+{1\over 15}\right ) .\nonumber\\
\label{amazing}
\end{eqnarray}
Nevertheless, we have not yet discovered a general formula for the
coefficients.
\bigskip
\bigskip

Discussions with M. S. Marinov and M. Moshe are kindly acknowledged. We thank
A. Gov-Ari and A. Menasen for assistance in some early numerical computations.
C.M.B. thanks the US Department of Energy for financial support for this
research. J.F. thanks the Department of Physics at Washington University for
its
hospitality. J.F. is supported by a Rothchild postdoctoral fellowship and
also in part by the Robert A. Welch Foundation and by NSF Grant PHY 9009850.
\pagebreak

\end{document}